\documentstyle[prl,aps]{revtex}
\begin{document}
\twocolumn
\title{Supersymmetric minisuperspace with non-vanishing fermion number}
\author{Andr\'as Csord\'as\thanks{Permanent address: Research Institute
for Solid State Physics, P.O. Box 49, H1525 Budapest, Hungary} and
Robert Graham}
\address{Fachbereich Physik, Universit\"at-Gesamthochschule Essen\\
45117 Essen\\ Germany}
\maketitle

\begin{abstract}
The Lagrangean of $N=1$ supergravity is dimensionally reduced to one
(time-like) dimension assuming spatial homogeneity of any Bianchi type
within class A of the classification of Ellis and McCallum. The
algebra of the supersymmetry generators, the Lorentz generators, the
diffeomorphism generators and the Hamiltonian generator is determined
and found to close. In contrast to earlier work, infinitely many
physical states with non-vanishing even fermion number are found to
exist in these models, indicating that minisuperspace models in
supergravity may be just as useful as in pure gravity.
\end{abstract}


\narrowtext
Minisuperspace models have long served as an enormeously useful testing
ground for new ideas in quantum gravity, ranging from explorations of
its mathematical structure to investigation in the many open problems
of quantum cosmology (see e.g. \cite{1,2,3}). More recently, also the
quantum theory of {\it supersymmetric} minisuperspace models has
attracted the interest of many authors (see e.g. \cite{4}-\cite{16})
for similar reasons. Our own interest was first aroused by the chaotic
classical nature of the Bianchi type IX models and the discovery
\cite{7} that supersymmetric versions have simple explicit analytical
solutions in the {\it empty} and {\it filled} fermion sectors which can
be interpreted as  wormhole states \cite{9} and, in other cases, as
Hartle-Hawking no-boundary states \cite{16}. However, it has been shown
meanwhile by a simple scaling argument, that these special solutions
found in the minisuperspace models have no direct counterpart in
4-dimensional supergravity because there states in the empty and filled
sectors cannot exist \cite{17} (for a dissenting opinion see
\cite{18}). This rather recent result makes appear disturbing another
result found some time ago \cite{8} and claimed as confirmed by several
groups of authors \cite{9,10,11,12}: namely that the solutions in the
empty and filled fermion sectors are the {\it only} solutions
(i.~e.~the only physical states satisfying all constraints) for all
supersymmetric minisuperspace models of Bianchi type in class A
\cite{19} (without matter-coupling and with the exception, with a
certain operator ordering \cite{11,12}, of Bianchi type I which is,
also for other reasons, very special in this class). An even stronger
result of this type was reported for anisotropic supersymmetric
minisuperspace models with a non-vanishing cosmological constant
\cite{13,14,15}: the present consensus in the literature is that
{\it no} physical states exist, in this case, at all. Taken together
these results force the conclusion that the physical states found in
the minisuperspace models have no counterpart in the full theory,
and vice-versa, which would render supersymmetric minisuperspace
models useless as models of full supergravity contrary to the situation
in pure gravity.

However, doubts may be raised. The quoted results for the
minisuperspace models are paradoxical from a general point of view: In
comparison to pure gravity, so it would appear from these findings,
these supersymmetric models are overconstrained for physically not yet
well understood reasons, even though supergravity certainly has
{\it more} physical degrees of freedom than gravity, not less, namely
the two additional physical degrees of freedom per space-point of the
Rarita-Schwinger field.

The present work has the purpose to clarify these issues for the
supersymmetric models of Bianchi type within class A of the
classification of Ellis and  McCallum \cite{19}. Here we shall restrict
ourselves to the case of a vanishing cosmological constant. We have
derived explicitely, in the metric representation, the dimensionally
reduced generators of supersymmetry transformations,
Lorentz transformations, and coordinate transformations within the
homogeneity group; we have then determined their closed graded algebra
explicitely. The form of this graded algebra allows us to determine the
form of the nontrivial physical states in all sectors with even fermion
number, which may take the values 0,2,4,6 in these models. In all of
these sectors nontrivial solutions are found to exist, of which
infinitely many have fermion number 2 and 4.
Comparing with the earlier work, where only states for fermion number
0 and 6 were found, we notice that there the Lorentz-invariant ansatz
for solutions in the two- und four-fermion sector was too special,
allowing in each sector for 2 Lorentz invariant amplitudes only, rather
than the 15 permitted ones. In our solutions all the 15
Lorentz-invariant amplitudes of the two-fermion sector appear
(similarly for the four-fermion sector) and are reexpressed in terms of
a single amplitude which must satisfy a Wheeler-DeWitt equation.
Therefore, the new type of solution we find to exist in the two- and
four-fermion sectors permits the free choice of the fermion sector,
corresponding to the choice of the initial state of the
Rarita-Schwinger field, and, in addition, just as much freedom in the
choice of initial conditions as the Wheeler-DeWitt equation of the
corresponding Bianchi models in pure gravity. Moreover, the new
physical states we find in the minisuperspace models are direct
analogues of physical states in {\it full} supergravity. As a
consequence, supersymmetric minisuperspace models recover their
significance as models of full supergravity.

We now proceed to present some more detail, however, omitting technical
points as much as possible. These we shall present in a full report of
this work elsewhere. We shall base our work on the Lagrangean of $N=1$
supergravity given in \cite{20} adopting all spinor conventions given
there. The excellent account of the Hamiltonian form of $N=1$
supergravity in the metric representation given in \cite{21,5} is also
freely used in the following. In the metric representation the
independent variables are taken to be the tetrad components ${e_p}^a$
(with Einstein indices $p=1,2,3$ from the middle of the alphabet and
Lorentz indices $a=0,1,2,3$ from the beginning of the alphabet) and
the Grassmannian components ${\psi_p}^\alpha$ (with spinor index
$\alpha=1,2$) of the Rarita-Schwinger field. The ${e_p}^a$ form the
metric tensor $h_{pq}={e_p}^ae_{qa}$ on the space-like homogeneity
3-surfaces in the symmetric basis of 1-forms $\bbox{\omega}^p$,
satisfying $d\bbox{\omega}^p =\frac{1}{2}(m^{pq}/h^{1/2})\varepsilon_{qrs}
\bbox{\omega}^r\otimes\bbox{\omega}^s$ in Bianchi type models of class
A. Here $h=\det h_{pq}$ and $\epsilon_{qrs}$ denotes the components of
the Levi-Civita tensor ({\it not} the tensor density). The constant
symmetric matrix $m^{pq}$ is fixed by the chosen Bianchi type
\cite{22}. It transforms as a tensor under all coordinate changes from
one {\it symmetric} basis to another one. Due to the choice of a
symmetric basis the ${e_p}^a$ and ${\psi_p}^\alpha$ are functions of
time only.

Starting from the supergravity Lagrangean in \cite{20}, a canonical
formulation of supergravity, restricted to the purely time-dependent
variables ${e_p}^a, {\psi_p}^\alpha$, can now be developed in the same
way as in \cite{21}. From the Lagrangean one defines, as usual, the
generalized momenta $\hat{p}^p_{\hphantom{p}a}$ and
$\hat{\pi}^p_{\hphantom{p}\alpha}$ of ${e_p}^a$ and ${\psi_p}^\alpha$,
respectively. The Poisson brackets must be replaced by Dirac brackets
due to the appearance and subsequent elimination of second class
constraints, by which the adjoint quantities ${\bar{\psi}_p}^{\hphantom
{p}\dot{\alpha}}$, ${\hat{\bar{\pi}}^p}_{\hphantom{p}\dot{\alpha}}$ are
eliminated. The Dirac brackets (and their Grassmannian generalizations
\cite{23}) are decoupled by the introduction of
${{p_+}^p}_a$, \cite{21,26} via
\begin{eqnarray}
\label{eq:3}
  {{\hat{\pi}^p}}_{\hphantom{p}\alpha} &=& 2{\pi^p}_\alpha\nonumber\\
  {{\hat{p}^p}}_{\hphantom{p}a} &=& {{p_+}^p}_a -{\textstyle\frac{V}
  {2}}h^{1/2}\varepsilon^{pqr}
    {\psi_q}^\alpha\sigma_{a\alpha\dot{\alpha}}
         {\cal C}_{rs}^{\dot{\alpha\beta}}{\pi^s}_\beta
\end{eqnarray}
with
\begin{equation}
\label{eq:4}
  {\cal C}_{pq}^{\dot{\alpha\alpha}} = -{\textstyle\frac{1}{2Vh^{1/2}}}
    \left(ih_{pq}n^a-\varepsilon_{pqr}e^{ra}\right)
       \bar{\sigma}_a^{\hphantom{a}\dot{\alpha}\alpha}\,.
\end{equation}
Here $\sigma_a$ are the $\sigma$-matrices
in the conventions of \cite{20}, $V$ is the volume of the compact
or compactified space-like homogeneity surface
$V=\int\bbox{\omega}^1\wedge\bbox{\omega}^2\wedge\bbox{\omega}^3$, and
$n^a$ is the future-oriented time-like unit vector orthogonal to the
space-like homogeneity surfaces. It must be kept in mind that $n^a$ is
a function of the tetrad components ${e_p}^b$. The only non-vanishing
Dirac brackets now are
$  \{{e_p}^a, {{p_+}^q}_b\}^* = {\delta_p}^q{\delta_b}^a$ and
$  \{{\pi_p}^\alpha, {\psi^q}_\beta\}^* =
      -{\delta_p}^q{\delta_\beta}^\alpha$.
The supersymmetry generators $S_\alpha$, $\bar{S}_{\dot{\alpha}}$ and
Lorentz generators $J_{\alpha\beta}$, $\bar{J}_{\dot{\alpha}\dot
{\beta}}$ in this representation are obtained as
\begin{eqnarray}
\label{eq:6}
  S_\alpha &=& -\left(\frac{1}{2}Vm^{pq}{e_q}^a+{\textstyle\frac{i}{2}}
                 {p_+}^{pa}\right)
   \sigma_{a\alpha\dot{\alpha}}
  {\cal C}_{pr}^{\dot{\alpha}\beta}{\pi^r}_\beta\nonumber\\
  \bar{S}_{\dot{\alpha}} &=& \left(\frac{1}{2}Vm^{pq}{e_q}^a
     -{\textstyle\frac{i}{2}}{p_+}^{pa}\right)
    \sigma_{a\alpha\dot{\alpha}}{\psi_p}^\alpha
\end{eqnarray}
and
\begin{eqnarray}
\label{eq:7}
  J_{\alpha\beta} &=& +{\textstyle\frac{1}{2}}
       (\sigma^{ac}\epsilon)_{\alpha\beta}
    \left(e_{pa}{{p_+}^p}_c-e_{pc}{{p_+}^p}_a\right)
  \nonumber\\
      && -{\textstyle\frac{1}{2}}
         \left(\psi_{p\alpha}{\pi^p}_\beta+\psi_{p\beta}
           {\pi^p}_\alpha\right)
  \nonumber\\
  \bar{J}_{\dot{\alpha}\dot{\beta}} &=& -{\textstyle\frac{1}{2}}
    (\epsilon\bar{\sigma}^{ac})_{\dot{\alpha}\dot{\beta}}
     \left(e_{pa}{{p_+}^p}_c-e_{pc}{{p_+}^p}_a\right)\,.
\end{eqnarray}
We have checked that the Dirac bracket
$  H_{\alpha\dot{\alpha}} = -2i\left\{S_\alpha\,,\,
    \bar{S}_{\dot{\alpha}}\right\}^*
$
differs only by terms proportional to $J_{\alpha\beta}$ or
$\bar{J}_{\dot{\alpha}\dot{\beta}}$ from the expression
$ \tilde{H}_{\alpha\dot{\alpha}} =\sigma_{a\alpha\dot{\alpha}}
   \left({e_p}^a{\cal H}^p+n^a{\cal H}\right)
$
defined by the generators of the Hamiltonian ${\cal H}$ and
diffeomorphism constraint ${\cal H}^p$ constructed in the canonical
formulation.

Canonical quantization is achieved by the requirements
$  {{p_+}^p}_a=-i\hbar(\partial /\partial{e_p}^a) $ and
$   {\pi^p}_\alpha = -i\hbar(\partial /\partial{\psi_p}^\alpha)$.
In $S_\alpha$ there is then an ordering ambiguity between ${{p_+}^p}_a$
and ${\cal C}_{pr}^{\dot{\alpha}\beta}$, which we have resolved by
adopting the ordering as written in
eq.~(\ref{eq:6}), see also \cite{21}. Using the quantized generators we
then have computed their algebra. The result of these calculations,
which are quite lengthy, in particular for establishing
eqs.~(\ref{eq:16}), (\ref{eq:17}), is as follows:
\begin{eqnarray}
\label{eq:9}
  \Big[S_\alpha,S_\beta\Big]_+ &=& 0
     = \Big[\bar{S}_{\dot{\alpha}},\bar{S}_{\dot{\beta}}\Big]_+ \\
\label{eq:10}
  \Big[S_\alpha,\bar{S}_{\dot{\alpha}}\Big]_+ &=& -\frac{\hbar}{2}
   H_{\alpha\dot{\alpha}}
\end{eqnarray}
The operator $H_{\alpha\dot{\alpha}}$ is here {\it defined} by the
anticommutator (\ref{eq:10}) in accordance with the classical Dirac
bracket. The commutation relations of all operators with $J_{\alpha
\beta}$ and $\bar{J}_{\dot{\alpha}\dot{\beta}}$ need not be written
down explicitely, as they simply reflect the transformation properties
of spinors and Lorentz-tensors in the spinor-formalism.
In particular $S^\alpha S_\alpha$ and $\bar{S}_{\dot{\alpha}}
\bar{S}^{\dot{\alpha}}$
are Lorentz scalars and therefore commute with $J_{\alpha\beta}$,
$\bar{\bar{J}_{\dot{\alpha}\dot{\beta}}}$.
\begin{eqnarray}
\label{eq:16}
  \Big[H_{\alpha\dot{\alpha}},\bar{S}_{\dot{\beta}}\Big]_- &=&
    i\hbar\varepsilon_{\dot{\alpha}\dot{\beta}}{D_\alpha}^{\beta\gamma}
    J_{\beta\gamma}\\
\label{eq:17}
   \Big[H_{\alpha\dot{\alpha}},S_{\beta}\Big]_- &=&
      -i\hbar\varepsilon_{\alpha\beta}\bar{J}_{\dot{\beta}\dot{\gamma}}
 {\bar{D}_{\dot{\alpha}}}^{\hphantom{\alpha}
              \dot{\beta}\dot{\gamma}}\nonumber\\
    &=& -i\hbar\varepsilon_{\alpha\beta}
     \Bigl[{\bar{D}_{\dot{\alpha}}}^{\hphantom{\alpha}\dot{\beta}\dot
     {\gamma}}\bar{J}_{\dot{\beta}\dot{\gamma}}+
        i\hbar{E_{\dot{\alpha}}}^{\gamma\delta}J_{\gamma\delta}
      \Bigr. \nonumber\\
    & & \hphantom{-i\hbar\varepsilon_{\alpha\beta}} \Bigl.
-i\hbar(1/Vh^{1/2})n^a\varepsilon_{\dot{\alpha}\dot{\beta}}
         {\bar{\sigma}_a}^{\hphantom{a}\dot{\beta}\gamma}S_\gamma
     \Bigr]
\end{eqnarray}
The commutators $[\bar{S}_{\dot{\beta}}$, $H_{\alpha\dot{\alpha}}]_-$
and $[S_{\beta}, H_{\alpha\dot{\alpha}}]_-$ are the essential new
results, on which all of the following is based. It should be noted, in
particular, that only Lorentz generators appear on the right hand side
of eq.~(\ref{eq:16}).  The operators ${D_\alpha}^{\beta\gamma}$,
${\bar{D}_{\dot{\alpha}}}^{\hphantom{\alpha}\dot{\beta}\dot{\gamma}}$,
${E_{\dot{\alpha}}}^{\beta\gamma}$ are odd and functions of
${\psi_p}^\alpha$, ${\pi^p}_\alpha$, ${e_p}^a$, ${{p_+}^p}_a$, whose
explicit form we have determined, but will not be important here. The
results (\ref{eq:9})-(\ref{eq:17}) demonstrate that the algebra of the
generator closes, a fact which was always assumed in the earlier work
on the same minisuperspace models, but which is here established
explicitely by eqs.~(\ref{eq:16}), (\ref{eq:17}).

Let us now find the physical states of these models, given by all the
states which are annihilated by the generators $S_\alpha$,
$\bar{S}_{\dot{\alpha}}$,
$J_{\alpha\beta}$, $\bar{J}_{\dot{\alpha}\dot{\beta}}$,
$H_{\alpha\dot{\alpha}}$.
The Lorentz generators automatically annihilate all states which are
Lorentz scalars. Therefore, it is sufficient to demand that physical
states are Lorentz scalars and annihilated by $S_\alpha$ and
$\bar{S}_{\dot{\alpha}}$; their annihilation by $H_{\alpha\dot
{\alpha}}$ is then automatically guaranteed due to eq.~(\ref{eq:10}).
The form of the constraint operators guarantees that physical states
have a fixed fermion number $F={\psi_p}^\alpha\partial/\partial
\psi^{\alpha}_p$, given by the number of factors of ${\psi_p}^\alpha$
in the $\psi$-representation: $F$ must be an even number in
Lorentz-invariant states and ranges from 0 to 6 in the present models.

The physical states in the sectors $F=0$ and $F=6$ are easily
obtained, and are, respectively, given by
\begin{eqnarray}
\label{eq:19}
  \Psi_0 &=& {\mbox{\rm const}\,} e^{\frac{V}{2\hbar}m^{pq}h_{pq}}
             \nonumber\\
  \Psi_6 &=& {\mbox{\rm const}\,h\,} e^{-\frac{V}{2\hbar}m^{pq}h_{pq}}
     \prod_r\left(\psi_r\right)^2
\end{eqnarray}
reproducing a well known result \cite{7}-\cite{12}.

In order to show that there exist physical states in the 2-fermion
sector let us consider the wave-function
\begin{equation}
\label{eq:20}
    \Psi_2 = \bar{S}_{\dot{\alpha}}\bar{S}^{\dot{\alpha}}f(h_{pq}),
\end{equation}
where we require,of course, that
$\bar{S}_{\dot{\alpha}}\bar{S}^{\dot{\alpha}}f \ne 0 $.
Here $f$ is a function of the $h_{pq}$ only, and therefore, like
$\bar{S}_{\dot{\alpha}}\bar{S}^{\dot{\alpha}}$, a Lorentz scalar.
Therefore $\Psi_2 $ {\it automatically} satisfies the Lorentz
constraints and the $\bar{S}$-constraints because of eq.~(\ref{eq:9}).
The only remaining condition is $S_\alpha\Psi_2 = 0$, which, after
the use of eqs.~(\ref{eq:20}), (\ref{eq:10}), reduces to
\begin{equation}
\label{eq:21}
  \Big[H_{\alpha\dot{\alpha}},\bar{S}^{\dot{\alpha}}\Big]_-f \quad
   +2\bar{S}^{\dot{\alpha}}H_{\alpha\dot{\alpha}}f \quad =0
\end{equation}
The first term is proportional to $J_{\beta\gamma}$ thanks to
eq.~(\ref{eq:16}) and therefore vanishes because $f$ is a Lorentz
scalar. The second term vanishes if $f$ satisfies the Wheeler-DeWitt
equation
\cite{A}
\begin{equation}
\label{eq:23}
 {H_{\alpha\dot{\alpha}}}^{(0)}f(h_{pq})=0
\end{equation}
where ${H_{\alpha\dot{\alpha}}}^{(0)}$ consists only of the bosonic
terms of $H_{\alpha\dot{\alpha}}$, i.e. of the terms which remain if
${\pi^p}_\alpha$ is brought to the {\it right} and then equated to
zero.
Any solution of this Wheeler-DeWitt equation, which may be specified
further, e.g. by imposing Hartle-Hawking \cite{2} no-boundary
conditions, or Vilenkin \cite{2} tunneling boundary conditions, or
wormhole boundary conditions \cite{24}, or some choice of scalar
product \cite{3}, generates a solution in the 2-fermion sector via
eq.~(\ref{eq:20}), with a definite dependence on the fermionic
variables, which are only present in the $\bar{S}_{\dot{\alpha}}
\bar{S}^{\dot{\alpha}} $-term.

Let us observe now that the norm of $\Psi_2$ vanishes due to the
appearance of $\bar{S}_{\dot{\alpha}}$ in eq.~(\ref{eq:20}) and the
fact that $S_\alpha$ is the adjoint of $\bar{S}_{\dot{\alpha}}$.
However, a proper definition of the scalar product, which we shall not
attempt here, must also include some gauge-fixing condition in its
measure. The norm of $\Psi_2$ (and of $\Psi_4$ to be considered below)
in such a properly defined scalar product will then not vanish.

Let us now turn to physical states in the 4-fermion sector. Similarly
to eq.~(\ref{eq:20}) the wave-function
\begin{equation}
\label{eq:24}
  \Psi_4 = S^\alpha S_\alpha g(h_{pq})\prod_{r=1}^3(\psi_r)^2\,.
\end{equation}
automatically satisfies the Lorentz constraints and the $S$-constraint.
It remains to satisfy the $\bar{S}_{\dot{\alpha}}$ constraint, which
reduces to
\begin{equation}
\label{eq:25}
 \left(\left[H_{\alpha\dot{\alpha}},S^\alpha\right]_-
        +2S^\alpha H_{\alpha\dot{\alpha}}\right)
         g(h_{pq})\prod_{r=1}^3(\psi_r)^2 = 0\,.
\end{equation}

Let us consider the first term in the bracket: By the use of equation
(\ref{eq:17}) it is expanded in terms containing the Lorentz generators
or $S_\gamma$ as factors on the right. The terms containing the Lorentz
generators vanish as they act on Lorentz scalars. In the term
containing $S_\gamma$, the generator $S_\gamma$ can be brought to the
left because it happens to commute with its prefactor. Combining this
term with the second term in the bracket of eq.~(\ref{eq:25}),
$S^\alpha$ can be factored out to the left. To fulfill
eq.~(\ref{eq:25}) it is therefore enough \cite{A} if $g$ satisfies the
Wheeler-DeWitt equation.
\begin{equation}
\label{eq:26}
  \Big({H_{\alpha\dot{\alpha}}}^{(1)}-\frac{\hbar^2}{Vh^{1/2}}
   n^a\sigma_{a\alpha\dot{\alpha}}\Big)g(h_{pq}) = 0
\end{equation}
where ${H_{\alpha\dot{\alpha}}}^{(1)}$ consists of those terms of
$H_{\alpha\dot{\alpha}}$ which remain if the ${\pi^p}_a$ are brought
to the {\it left} and then equated to zero. The resulting
Wheeler-DeWitt equation is slightly different from that obeyed by the
amplitude in the 2-fermion sector, but apart from this the degree of
generality of the solution is the same.

Our results differ from earlier work on the supersymmetric Bianchi
models in class A \cite{8}-\cite{12} which concluded that physical
states in the 2- and 4-fermion sectors do not exist. In hindsight, this
conclusion can be traced to an overly restricted ansatz for
Lorentz-invariant wave-functions \cite{25}:
Lorentz invariants were constructed only from the irreducible spin-1/2
and spin-3/2 components contained in the Rarita-Schwinger field
${\psi_p}^\alpha$ (there are only 2 such invariants bilinear and two
more quadri-linear in ${\psi_p}^\alpha$), omitting the further
invariants which can be formed with help of the irreducible spin-2
components contained in the gravitational degrees of freedom. In fact,
such components can be used to generate up to ${6\choose 2}=15$ Lorentz
invariants in the 2-fermion sector. A simple example of such
an additional invariant is $m^{pq}{\psi_p}^\alpha\psi_{q\alpha}$
another one is $m^{pr}h_{rs}m^{sq}{\psi_p}^\alpha\psi_{q\alpha}$ etc.
Writing out the expressions (\ref{eq:20}), (\ref{eq:23}) for $\Psi_2$
and $\Psi_4$ in an explicit way it can be seen that indeed they contain
such additional invariants.

We finish by speculating on a possible generalization of these
solutions for the case of full supergravity. Provided the algebra
of the local generators of the constraints still has a form like
eqs.~(\ref{eq:10})-(\ref{eq:17}) physical states still exist which look
somewhat like $\Psi_2$ and $\Psi_4$, but contain formal products of
$(\bar{S})^2$ or $(S)^2$ over all points of the space-like 3-surface,
thus leading to states with infinite fermion number. Such states have
recently been discussed in \cite{17}. The new physical states we have
found in the present paper are the direct minisuperspace analogues of
such states in full supergravity; even though, due to the reduction to
minisuperspace the fermion number is, of course, finite (in fact just 2
or 4). It is remarkable that the new states display the same richness
of gravitational dynamical behavior as the Bianchi models in pure
gravity. In this respect they differ qualitatively from the earlier
found states in the empty and the full fermion sectors, which are
highly symmetric and do not describe the full dynamical behavior of
the Bianchi models in the classical limit, which, as is well known, can
be very asymmetric and rich.

While the states in the empty and filled sectors alone would span at
most a 2-dimensional Hilbert space, the new physical states described
here span an infinite-dimensional Hilbert space just as in the Bianchi
models of pure gravity \cite{3}. Just how this Hilbert space ought to
be constructed is, of course, one of the questions one hopes to unravel
by the further study of {minisuperspace} models. That this strategy is
not only fruitful in the case of pure gravity but can also be usefully
employed in the case of {\it supergravity} is an encouraging conclusion
of the present paper.

This work has been supported by the Deutsche Forschungsgemeinschaft
through the Sonderforschungsbereich 237 ``Unordnung und gro{\ss}e
Fluktuationen''. One of us (A.~Csord\'as) would like to acknowledge
additional support by The Hungarian National Scientific Research
Foundation under Grant number F4472.


\begin{references}
\bibitem{A}  Whether the even more general {possi\nolinebreak bility}
             ${H_{\alpha\dot{\alpha}}}^{(0)}f(h_{pq})=C(n^a/h^{1/2})
             \sigma_{a\alpha\dot{\alpha}}\exp(Vm^{pq}h_{pq}/2\hbar)$
              to satisfy (11), with arbitrary constant $C\neq 0$ (and
              a similar inhomogenous generalization of eq.~(15), with
              $cn^a h^{1/2}\sigma_{a\alpha\dot{\alpha}}
              \exp(-Vm^{pq}h_{pq}/2\hbar)$
              appearing in the
              inhomogeneity) are relevant, too, can only be decided
              when the full Hilbert space of physical states is
              constructed. Here we take $C=0$.
\bibitem{25} In Ref. \cite{10} solutions of the form (\ref{eq:20})
             were actually looked for,
             but were there found not to exist, because in \cite{10}
             the generators and their algebra were not derived from
             supergravity, and different from eqs.
             (\ref{eq:9})-(\ref{eq:17}).
\end{references}
\end{document}